\documentclass[aps,prd,singlecolumn,tightenlines,superscriptaddress,nofootinbib,floatfix]{revtex4-2}

\usepackage{upgreek}

\usepackage[official]{eurosym}
\usepackage[normalem]{ulem}
\usepackage{amstext}
\usepackage{graphicx}
\graphicspath{{figures/}}
\usepackage{url}
\usepackage{color}
\usepackage{ulem}
\usepackage[version=4]{mhchem}
\usepackage[utf8]{inputenc}
\usepackage{fontawesome}
\usepackage{yfonts}

\pdfoutput=1
\usepackage{textcomp}
\usepackage{comment}
\usepackage{yfonts}
\usepackage{epsfig,amsfonts,mathrsfs,graphicx,color,slashed,multirow}
\usepackage{latexsym,graphicx,slashed,color,enumerate,url,cancel,gensymb}
\usepackage{textcomp}

\usepackage[x11names]{xcolor}
\usepackage[colorlinks,pdfstartview=FitV,breaklinks=true]{hyperref}
\usepackage{booktabs}
\usepackage{adjustbox}
\usepackage{textgreek} 
\usepackage{caption, subcaption} 
\usepackage{float}

\usepackage{lmodern}
\usepackage{ae,aecompl}
\usepackage{appendix}


\begin{document}


\title{
Multimessenger Concordance for the Cygnus Region as the Source of the Cosmic-Ray Knee 
}

\author{Luis E.~Espinosa Castro}
\email{luis.espinosacastro@gssi.it}
\affiliation{Gran Sasso Science Institute, Viale Francesco Crispi 7, 67100, L’Aquila, Italy}
\affiliation{INFN, Laboratori Nazionali del Gran Sasso, Via G. Acitelli 22, 67100, Assergi, Italy}
\author{Kotha Murase}
\email{murase@psu.edu}
\affiliation{Department of Physics, The Pennsylvania State University, University Park, Pennsylvania 16802, USA}
\affiliation{Department of Astronomy and Astrophysics, The Pennsylvania State University, University Park, Pennsylvania 16802, USA}
\affiliation{Institute for Gravitation and the Cosmos, The Pennsylvania State University, University Park, Pennsylvania 16802, USA}
\affiliation{Center for Gravitational Physics and Quantum Information, Yukawa Institute for Theoretical Physics, Kyoto, Kyoto 606-8502 Japan}

\author{Carlo Rizza}
\affiliation{Department of Physical and Chemical Sciences,
University of L’Aquila, 67100 L’Aquila, Italy}

\author{Francesco L.~Villante}
\affiliation{INFN, Laboratori Nazionali del Gran Sasso, Via G. Acitelli 22, 67100, Assergi, Italy}
\affiliation{Department of Physical and Chemical Sciences,
University of L’Aquila, 67100 L’Aquila, Italy}
\author{Vittoria Vecchiotti}
\affiliation{INAF Osservatorio Astrofisico di Arcetri Largo Enrico Fermi,5, 50125, Firenze, Italy}
\affiliation{Tsung-Dao Lee Institute, Shanghai Jiao Tong University, Shanghai 201210, P. R. China}
\author{Giulia Pagliaroli}
\email{giulia.pagliaroli@lngs.infn.it}
\affiliation{INFN, Laboratori Nazionali del Gran Sasso, Via G. Acitelli 22, 67100, Assergi, Italy}

\date{\today}

\begin{abstract}
The origin of the cosmic-ray (CR) knee remains one of the central open questions in particle astrophysics. Recent measurements by the Large High Altitude Air Shower Observatory revealed a pronounced feature in the proton spectrum at $\sim3-4$~PeV, while observations of diffuse gamma rays above $100$~TeV do not exhibit a corresponding spectral break. This apparent discrepancy challenges the standard interpretation, in which the local CR distribution is representative of the Galactic CR sea. Here, we investigate whether the CR knee can instead originate from the Cygnus region as a nearby PeVatron. By combining CR measurements at Earth with very-high-energy gamma-ray observations from LHAASO and the Tibet-AS$\gamma$ experiment, we identify an additional hard gamma-ray component in the inner Galaxy consistent with a source located in the Cygnus region. We show that our results provide a concordance multimessenger picture. The required properties are compatible with the PeVatron candidate detected by LHAASO in the Cygnus bubble and with the Galactic neutrino flux observed by the IceCube Neutrino Observatory. 
\end{abstract}

\maketitle

\section{Introduction}
\label{sec:intro}
Recent multimessenger observations have changed our view of very high-energy Galactic emission. Dozens of very high-energy gamma-ray sources in the $\sim0.1-1$~PeV range have been discovered by the High Altitude Water Cherenkov (HAWC) experiment and the Large High Altitude Air Shower Observatory (LHAASO)~\cite{Abeysekara:2017hyn,LHAASO:2021gok}. The Galactic diffuse gamma-ray emission in the sub-PeV energy has also been measured by the Tibet AS$\gamma$-ray experiment \citep{Amenomori_2021} and LHAASO \citep{Cao_2025a}, which turned out to be comparable to the Galactic diffuse neutrino flux reported by the IceCube Collaboration \citep{IceCube:2023ame}. 
On the cosmic-ray (CR) side, a spectral break so-called the ``knee'' at $\sim3-4$~PeV is one of the most prominent features in the Galactic CR spectrum, yet its origin remains unresolved despite decades of observations. For a long time, it has been attributed to the maximum energy achievable by Galactic accelerators (e.g. supernova remnants) or by the change in propagation effects and CR confinement in the Galaxy \citep{Ptuskin_1993, Giacinti_2015}. However, recent measurements extending into the PeV regime have provided a unique opportunity to test these hypotheses or other alternative models. In particular, measurements of the proton spectrum by LHAASO have revealed a prominent feature at energies of $\sim3-4$~PeV, corresponding to the well-known CR knee \citep{LHAASO:2025byy}. The CR knee could also be attributed to the transition towards another class of accelerators \citep{Sveshnikova:2003sa,Murase:2013kda,Zhang:2025tew,Kaci2025} or to the presence of nearby sources \citep{Prevotat_2025,Fang:2026ydz}. Indeed, direct and ground-based measurements of the CR flux alone seem to indicate that Galactic CRs are composed of distinctive populations originated by different classes of accelerators \citep{Vaiman:2025duv,Aharonian2026}. Moreover, recent studies have shown that the discreteness of Galactic sources can produce sizable fluctuations in diffuse gamma-ray and neutrino emission, which also motivates regional interpretations of the high-energy sky that go beyond smooth background models \citep{Stall:2025ggd}.

Surprisingly, a corresponding spectral feature is not observed in the observations of the Galactic diffuse gamma-ray emission above 100~TeV by LHAASO \citep{Cao_2025a}. This is because the spectrum feature around $E_p\sim3-4$~PeV leads to a corresponding feature at $E_\gamma \sim 0.08E_p\sim(200-300)$~TeV (e.g., Ref.~\cite{Ahlers:2013xia}). This apparent tension was quantified assuming that the diffuse gamma-ray emission arises from interactions of CRs with the interstellar medium \citep{Castro:2025wgf}. The absence of a gamma-ray counterpart to the CR knee suggests that additional components may contribute to the local CR spectrum. 

\begin{figure}[t!]
\centering
\includegraphics[scale=0.5]{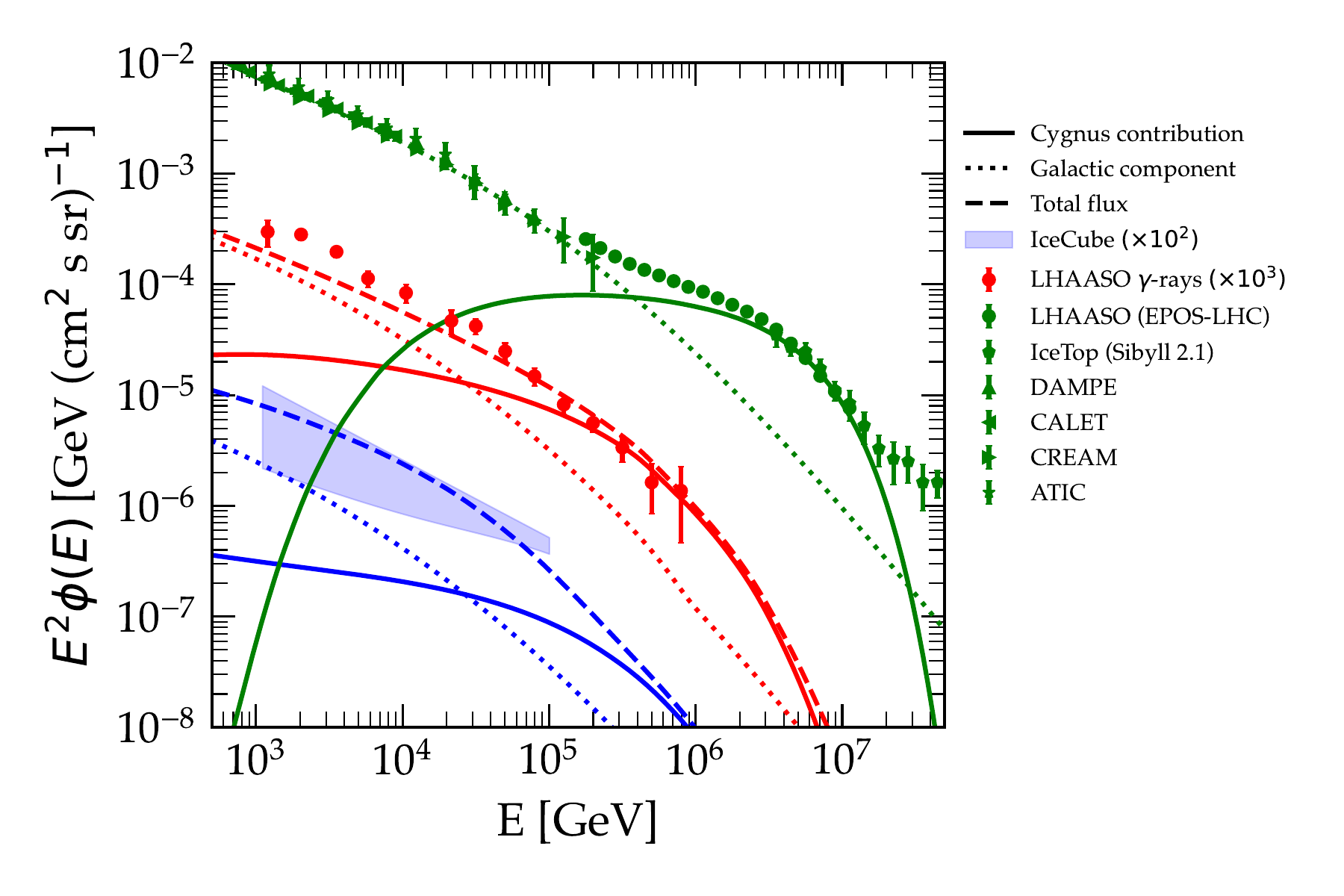}
\caption{Summary plot of our results on contributions from the Cygnus region to the CR proton flux (solid green line), diffuse gamma-ray and neutrino fluxes (red and blue solid lines, respectively), as observed in different sky regions. 
The Galactic CR sea component is shown by the dotted green line and its diffuse gamma-ray and neutrino components are shown with red and blue dotted lines, respectively. Total gamma-ray and neutrino fluxes (corresponding to the sum of the Galactic and Cygnus components, as well as the contribution from unresolved sources) are displayed with red and blue dashed lines, respectively. For details on the computation of each flux component, see Sections \ref{sec:Methods} and \ref{sec:Results}. 
For gamma rays, the flux is computed for the LHAASO inner Galactic region ($|b|<5^\circ$, $15^\circ<l<125^\circ$); while for neutrinos, the flux is computed for the Galactic plane region of IceCube ($|b|<90^\circ$, $0^\circ<l<360^\circ$). Cygnus proton fluxes are compared to observations by direct missions (DAMPE \citep{An_2019}, CALET \citep{Adriani_2019}, CREAM \citep{Yoon_2017} and ATIC-2 \cite{Panov_2009}) and ground-bases experiments (IceTop \citep{Aartsen_2019} and LHAASO \citep{Cao_2025b, LHAASO2025}), shown with green scatter points. Observations of diffuse gamma rays in the corresponding LHAASO sky region \citep{Cao_2025a} displayed as red scatter points. Best-fit Galactic diffuse (all-flavor) neutrino flux based on IceCube observations \citep{IceCube:2023ame} for three different templates ($\pi^0$ and KRA$^5$ and KRA$^{50}$) 
are shown as blue shaded area, spanned by the minimum and maximum best-fit fluxes.}
\label{fig:Multi_Messenger}
\end{figure}

In this work, we explore a multimessenger scenario, in which the CR knee receives a substantial contribution from a nearby PeVatron in the Cygnus region. This region includes a young massive OB association (Cygnus OB2), which is hosted by the Cygnus cocoon with a size of $\sim50-70$~pc and Cygnus bubble with $\sim150$~pc, which are observed by Fermi-LAT \cite{Fermi:2011} and LHAASO \cite{LHAASO:2023uhj}, respectively, as well as several plausible Galactic PeVatron candidates, such as Cygnus X-1. IceCube observations of the Galactic plane may also hint at a possible neutrino excess broadly compatible with this region \cite{IceCube:2023ame,Neronov:2023hzu}. Motivated by these observations of the Cygnus region, through the combination of CR measurements at Earth with diffuse gamma-ray observations from LHAASO and Tibet AS\(\gamma\) experiment, we constrain the source properties and derive the associated CR, gamma-ray, and neutrino signals. We show that the Cygnus-origin knee component can provide a consistent multimessenger interpretation of the available data within current uncertainties (see Fig.~\ref{fig:Multi_Messenger}). We find that the unmasked diffuse flux measured by LHAASO in the inner Galaxy exceeds that in the outer Galaxy above $E_\gamma \gtrsim 100$~TeV, corresponding to the knee energy. Consistently, the Tibet AS\(\gamma\) data, once rescaled by gas column density, favor a source contribution from a longitude range of $70^\circ \lesssim l \lesssim 90^\circ$.

\section{Methods}
\label{sec:Methods}

\subsection{The Cygnus region as a CR reservoir embedding a PeVatron}
\label{Sec:Pevatron}
\begin{figure}[t]
\centering
\includegraphics[scale=0.2]{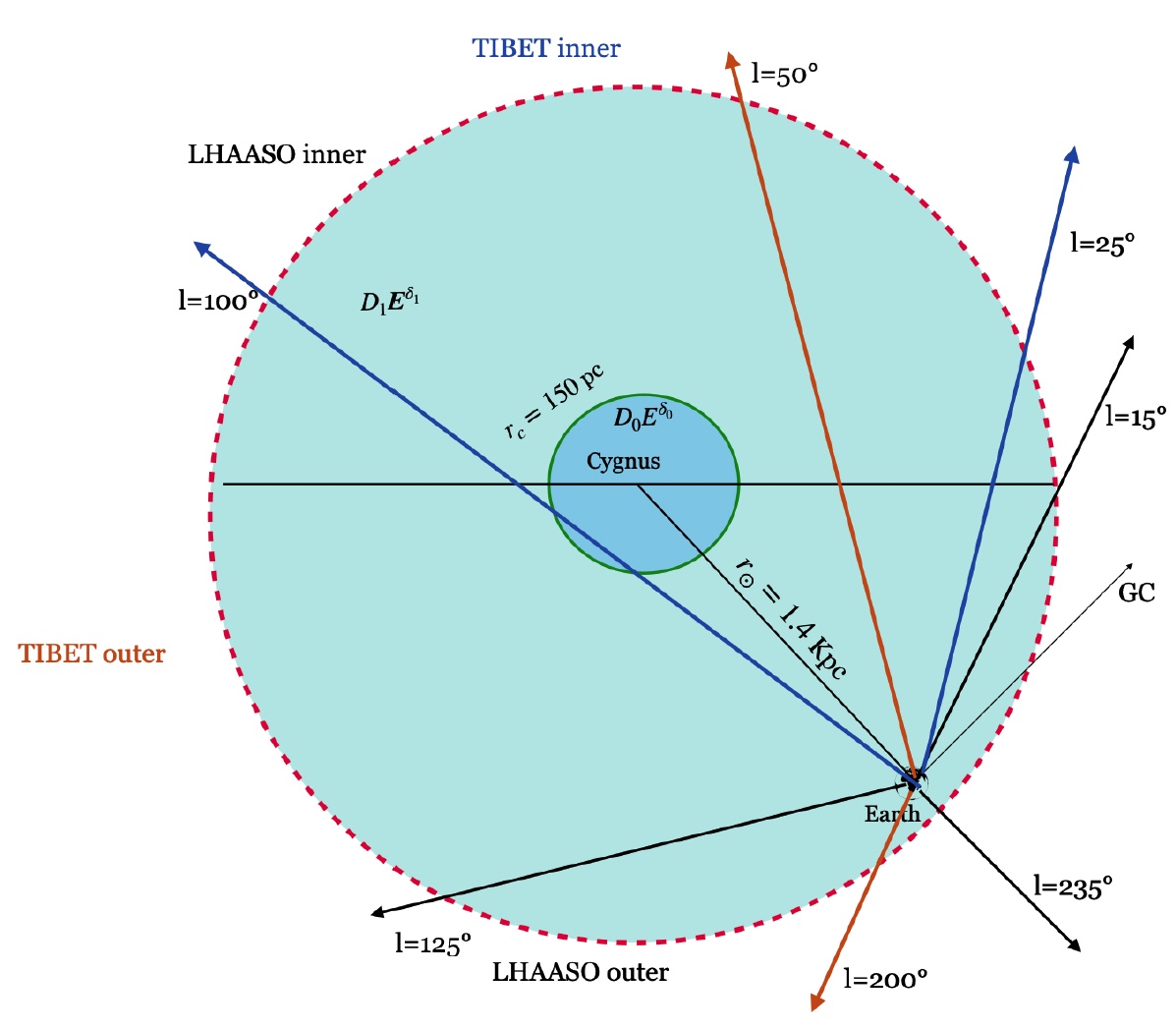}
\caption{Schematic representation (not in scale) of the Cygnus region, as located in the context of the two LHAASO and Tibet-AS$\gamma$ regions and with respect to the Earth position. With a green circle we illustrate the slow diffusion zone (the Cygnus bubble) and with a dashed red lines we show the diffusion region of PeV protons.}
\label{fig:PeVatron}
\end{figure}

We assume that a PeVatron exists inside the Cygnus region. Such a source may naturally exist inside the Cygnus bubble at a distance $d_{\rm Cyg}=1.4$~kpc from the Earth, with Galactic coordinates of $(l,b)=(79.61^\circ,1.65^\circ)$. A schematic representation of this scenario is reported in Fig.\ref{fig:PeVatron}. 

 The Cygnus bubble or its inner region, Cygnus cocoon, hosts Cygnus OB2, which has a total stellar mass of $M_\ast \approx 1.7\times 10^4\,M_\odot$ and a dominant star-formation episode lasting over $\sim 1-7$~Myr \cite{Wright:2015}, implying an average star-formation rate of $\sim 3\times 10^{-3}\,M_\odot\,{\rm yr^{-1}}$. Its kinetic power is estimated to be $L_{\rm kin}\sim3\times 10^{38}\,{\rm erg\,s^{-1}}$, indicating that the cumulative CR energy input can be $W_{\rm CR}\approx\epsilon_{\rm CR}L_{\rm kin} t_{\rm age}\sim 2\times {10}^{51}~{\rm erg}~\epsilon_{\rm CR,-1}L_{\rm kin,38.5}(t_{\rm age}/6~{\rm Myr})$. For the bigger Cygnus region, the OB-star census is comparable to, or somewhat larger than, that of Cygnus OB2 alone. The maximum proton energy in a superbubble in a collective-wind shock scenario may be written as $E_{\rm max}^{\rm SB}\approx (3/20) Ze B R (V_s/c)\simeq2.8~{\rm PeV}~Z(B/100~\mu{\rm G})(R/20~{\rm pc})(V/3000~{\rm km}~{\rm s}^{-1})$. However, while CRs could be accelerated at such a termination shock \cite{Menchiari:2024tqw}, this should be regarded as an idealized estimate, and detailed hydrodynamical simulations indicate that Cygnus OB2 is too spatially extended to sustain a single large-scale collective wind shock, contrary to young massive star clusters \cite{Vieu:2024qjx,Haerer:2025ull}. Alternatively, a PeVatron can be a powerful supernova remnant such as a hypernova remnant \cite{Sveshnikova:2003sa,Ahlers:2013xia,Fang:2021ylv,Haerer:2025ull} (see also Ref.~\cite{Kimura:2013}), for which a single event has a kinetic energy of ${\mathcal E}_{\rm kin}\sim10^{52}\,\mathrm{erg}$, and the maximum energy is $E_{\rm max}^{\rm HN}\approx (3/20) Ze B R (V_s/c)\simeq3.1~{\rm PeV}~Z\epsilon_{B,-3}^{1/2}{\mathcal E}_{\rm kin,52}{(M_{\rm ej}/10~M_\odot)}^{-2/3}n_{\rm ISM}^{1/6}$ around the Sedov time \cite{Ahlers:2013xia}, where $\epsilon_B$ is the energy fraction carried by magnetic fields, $M_{\rm ej}$ is the ejecta mass, and $n_{\rm ISM}$ is the interstellar material density. The sustained star-formation activity of the Cygnus region also allows us to expect core-collapse supernovae on timescales of roughly $\sim10^4~\mathrm{yr}$, so over a few Myr the cumulative input from hundreds of supernova remnants, which may include a few hypernova remnants, can naturally reach $W_{\rm CR}\sim10^{51}-10^{52}\,\mathrm{erg}$.

In this work, we assume that a PeVatron exists in the Cygnus region without specifying details of the acceleration mechanism, and the proton injection spectrum is phenomenologically modeled as:
\begin{equation}
\dot{Q}(E_p)=\dot{Q}_0\left(\frac{E_p}{E_0}\right)^{-s} \exp{\left(-\frac{E_p}{E_p^{\rm max}}\right)}
\end{equation}
where $E_0=1$ TeV and $\dot{Q}_0$ is related to the total proton injection luminosity $L_p$ through $L_p=\int dE_p \, E_p Q(E_p)\, $.
We adopt a total proton energy of $W_p=2.8\times10^{51}$~erg with a spectral index of $s=2.1$ and a maximum proton energy of $E_p^{\rm max}=5$~PeV. Hereafter, for the purpose of demonstration, we adopt a proton luminosity of $L_p=3\times10^{37}$~erg/s that is continuously injected over $t_{\rm age}=3$~Myr, consistent with the injection parameters reported in Ref.~\cite{LHAASO:2023uhj}. 
Here we note that there are some degeneracies between the superbubble (continuous injection) and hypernova remnants (bursty injection episodes) interpretations (see also Ref.~\cite{Fang:2021ylv}). We also note that although Cygnus X-1, which is a microquasar at $d=2.2$~kpc, is a promising PeVatron, the inferred luminosity may be too small to satisfy the energetics constraint. 
%

The Cygnus bubble, which is seen by LHAASO, extends up to an angular radius of $\sim 6^\circ$, corresponding to a physical size $r_c\sim150$ pc. Within this region, the observed gamma-ray emission shows an angular distribution compatible with a proton density profile scaling as $\propto 1/r$ toward the center \cite{Aharonian:2018oau}. 
Such a behavior is expected in the case of homogeneous three-dimensional diffusion with continuous particle injection. 
However, we note that this $1/r$ behavior does not correspond to a direct fit to the observed gamma-ray surface brightness, given the asymmetric morphology and the probable multicomponent structure of the Cygnus bubble (cf. Ref.~\cite{Kimura:2013} based on X-ray observations). 

The gamma-ray emission from the Cygnus bubble depends on both the ambient gas density and the CR density. Close to the source, the observed gamma-ray emission is consistent with a diffusion coefficient in the form, $D=D_0(E_p/E_0)^{\delta_0}$ where $D_0\sim 3\times 10^{26} ~\text{cm}^2/\text{s}$, $E_0=1$~TeV, and $\delta_0=0.7$ \cite{LHAASO:2023uhj}.
This is about $2-3$ orders of magnitude smaller than the standard value of the diffusion coefficient in the interstellar medium. The Cygnus region is regarded as a reservoir of cosmic rays in the sense that CRs supplied from a PeVatron are confined during the diffusion time.  

To account for this suppressed-diffusion in the bubble, we adopt a spherically symmetric two-zone diffusion model described by the following diffusion equation
\begin{equation}
\label{eq:diffeq}
\frac{\partial n}{\partial t} = \frac{1}{r^2} \frac{\partial}{\partial r} \left( D  r^2  \, \frac{\partial n}{\partial r} \right)  + Q,
\end{equation}
where $n(r,t, E_p)$ is the density of protons at a radial distance from the source $r$ and time $t$ for a given proton energy $E_p$, the diffusion coefficient $D(r, E_p)$ is given by:
\begin{equation}
\label{eq:diffcoeff}
D(r, E_p) = \begin{cases}
 D_0(E_p/E_0)^{\delta_0} & 0 < r < r_c, \\
 D_1(E_p/E_0)^{\delta_1} & r \geq r_c, \\
\end{cases}  
\end{equation}
and $Q$ is an injection term.  In Eq.~(\ref{eq:diffcoeff}), the inner region $r<r_c$ is characterized by the suppressed diffusion, consistent with LHAASO gamma-ray observations \cite{LHAASO:2025byy}, whereas, in the outer region $r\geq r_c$, we assume Kolmogorov diffusion with $D_1=10^{28}~\text{cm}^2/\text{s}$ and $\delta_1=1/3$. 
%
To obtain an exact solution of Eq.~(\ref{eq:diffeq}), we solve it numerically for an injection term,
\begin{equation}
Q(r,t,E_p) = \dot{Q}(E_p)  G(r) \theta(t),
\end{equation}
with $G(r) = \frac{1}{(2\pi R^2)^{3/2}} \exp\left(-\frac{r^2}{2 R^2}\right)$, a Gaussian distribution characterized by a physical size of a PeVatron, $R=1$~pc, and $\theta(t)$ is the Heaviside step function, which describes a proton injection starting at $t=0$ and continuing up to the present time corresponding to the source age $t_{\rm age}$.  

We finally discuss the relevance of energy losses due to inelastic $pp$ interactions. Assuming an average interstellar medium number density $n_{\rm ISM} \sim1 \,\rm cm^{-3}$, the timescale for $pp$ interactions is $t_{\rm pp}=(c\, n_{\rm ISM}\, \sigma_{pp})^{-1}\sim10^{7}~\rm yr$.
This can be compared with the source age and the diffusion timescale $t_{\rm diff}$ over the source-Earth distance $d$, given by $t_{\rm diff}=d^2/[6D_1(E_{\rm p}/E_0)^{\delta_1}]+r_c^2/[6D_0(E_{\rm p}/E_0)^{\delta_0})]$. 
Even if we take the suppressed diffusion coefficient around the source, 
the resulting diffusion time scale is shorter than the interaction time scale for proton energies larger than $E_{p}\gtrsim 100$~TeV.
Therefore, at high energies, energy losses due to $pp$ interactions can be safely neglected.

\subsection{The CR flux and spectrum}
\label{sec:CR}
We evaluate the Cygnus PeVatron contribution to the CR density across the Galaxy by solving the diffusion equation for different source ages $t_{\rm age}$. The proton flux produced by the Cygnus region at the Earth is then calculated as $\phi_p^{\text{Cyg}}(E_p, t_{\rm age}) = (c/4\pi) \, n(E_p, r=d_{\rm Cyg} ,t_{\rm age})$, where $c$ is the speed of light and $d_{\rm Cyg}=1.4$ kpc is distance to the Cygnus region.

%
%
%

As discussed in Sec. \ref{sec:Results}, our results suggest that the Cygnus PeVatron is likely a significant contributor to the locally observed CR proton knee. This also affects the determination of the "CR sea" produced by sub-PeV Galactic sources, specifically their softening at few hundreds of TeV. Once the Cygnus component is fixed, the large-scale Galactic proton spectrum $\phi^{\text{Gal}}_{p}(E_{p})$ is indeed obtained by requiring that the sum of the Cygnus and Galactic contributions reproduces the experimental data.
\par The Galactic component is parameterized as a broken-power law \cite{Antonyan_2000} of the form,
\begin{equation}
  \label{eq:galactic_cr_flux}
  \phi^{\text{Gal}}_{p}(E_p) = N \left(\frac{E_p}{\bar{E}}\right)^{-\alpha_{1}}  \Pi_{i}\left[1+\left(\frac{E_p}{E_{b,i}}\right)^\frac{1}{\omega_i}\right]^{-\Delta \alpha_{i} \, \omega_i}
\end{equation}
where $N$ is a normalization factor with units (GeV cm$^2$ s sr)$^{-1}$ and $\bar{E}=10^2$~GeV. Eq.~\ref{eq:galactic_cr_flux} models the spectrum as a series of breaks located at energies $E_{b,i}$ with spectral index changing by $\Delta \alpha_i = \alpha_{i+1} - \alpha_{i}$ over an energy-width $\omega_i$ in logarithmic scale. We progressively add spectral breaks until the fit no longer improves and we find that three breaks (at rigidities of about 0.5, 10 and 300 TV) better reproduce the experimental data in all scenarios. 

If the Cygnus source also accelerates $^{4}{\rm He}$ nuclei, then it may significantly contribute to the local CR $^{4}{\rm He}$ knee (which is currently also measured by LHAASO \citep{LHAASO2025}). We calculate this component by assuming that the $^{4}{\rm He}$ injection spectrum and diffusion coefficients scale with rigidity with respect to the proton case.
%
%
As it is discussed, e.g., in Ref.~\cite{Castro:2025wgf}, the Cygnus $^{4}{\rm He}$ component is not expected to produce a large contribution to the gamma-ray and neutrino production at PeV energies because, due to rigidity scaling, it has a cutoff at an energy per nucleon which is a factor 1/2 smaller than protons. Nevertheless, it can have a relevant role for the interpretation of local CR data. Thus, we include it by fitting the large-scale Galactic $^{4}{\rm He}$ flux to observational data, using the same parametrization as for protons and assuming that spectral breaks are rigidity dependent, see Eq.~(\ref{eq:galactic_cr_flux}).

Once the Galactic $^{4}{\rm He}$ contribution is known, the Galactic heavy component is determined by following the approach described in Ref.~\cite{Castro:2025wgf}, i.e. by assuming heavy elements share the same rigidity dependence as helium, differing only by a normalization factor. From an energetics point of view, the required CR proton energy, $W_{p}$, is compatible with gradual energy injection by a long-lived superbubble powered by stellar winds, but it can also accommodate an additional impulsive injections from one or a few exceptionally energetic explosions such as hypernovae, occurring within the same environment. The present data therefore support the existence of a powerful and long-lived CR reservoir in the Cygnus region, while leaving some degeneracy in the detailed nature of the underlying accelerator.

\subsection{The Galactic diffuse gamma-ray and neutrino fluxes and spectra}
\label{sec:diffuse}
The diffuse gamma-ray and (one-flavor) neutrino fluxes produced by CR interactions with the interstellar medium can be written as~\cite{Pagliaroli:2016, Cataldo:2019qnz}:
\begin{equation}
\label{eq:gamma_flux}
\phi_{i} (E_i,\hat{n}_i)=
C_i \int_{E_\gamma}^{\infty} d E_{\rm n} \,
\frac{d\sigma_i}{dE_i}\left(E_{\rm n},\, E_i \right)\,
\int_{0}^{\infty} ds \,\phi_{\rm nuc} (E_{\rm n},{\bf r}_{\tiny \odot}+s \, \hat{n}_i )\,
n_{g} ({\bf r}_{\tiny \odot}+s \, \hat{n}_i)\, e^{-\tau_i(s, E_i)}
\end{equation}
where $i = \nu,\, \gamma$ stands for neutrinos and gamma-rays respectively, while $E_i$ and $\hat{n}_i$ indicate the energy and arrival direction of the considered particles.
The constant $C_i$ is equal to 1 for photons and 1/3 for neutrinos to take into account approximate flavor equipartition due to the neutrino mixing. 

%

%
The function $\phi_{\rm nuc} (E_{\rm n},{\bf r})$ represents the total CR
nucleon flux given by
\begin{equation}
\phi_{\rm nuc} (E_{\rm n}, {\bf r}) = \sum_A  A^2 \phi_{A} (A E_{\rm n}, {\bf r})
\end{equation}
where $A$ indicates the mass number of different nuclear species, $E_{n}$ is the energy per nucleon and ${\bf r}$ indicates the position in the Galaxy (with ${\bf r}_{\odot}$ denoting the position of the Sun).
For the proton component, we include both the Cygnus PeVatron contribution and the large-scale Galactic component which is assumed to be spatially uniform, i.e.,
\begin{equation}
\phi_{p} (E_{\rm n}, {\bf r_{\odot}}) =  \phi^{\rm Cyg}_{p} (E_{\rm n}, {\bf
  r_{\odot}}) + \phi^{\rm Gal}_{p} (E_{\rm n}).
\end{equation}
As it is explained in the previous sections, the relative weight  of
these components strongly depends on the distance from the source, leading to effects dependent on the observation directions that will be discussed in the following. 

The CR nucleon flux is convolved with the differential cross section $d\sigma_i/dE_i$ for gamma-ray and (all flavours) neutrino production in nucleon-nucleon collisions. 
We use the AAFRAG parameterization~\cite{Koldobskiy_2021, Kachelriess:2022khq} for the photon and neutrino production cross sections, in closest agreement with the approach of Ref.~\cite{Orusa:2023azt}, based entirely on accelerator data fits. 
We include gamma-ray absorption that becomes relevant at PeV energies by calculating the photon optical depth $\tau_\gamma(s,E_\gamma)$ due to pair production on background radiation field photons (mainly the cosmic microwave background), based on the parametrization in Ref.~\cite{Vernetto:2016alq}. Neutrinos free stream without being absorbed; thus the exponential factor in the r.h.s of Eq.~(\ref{eq:gamma_flux}) can be omitted, or equivalently $\tau_\nu(s, E_\nu) \equiv 0$ can be assumed.

The function $n_{g}({\bf r})$ represents the interstellar gas density.
This is mainly composed of atomic (H) and molecular hydrogen (H$_2$), whose distributions are traced by the 21~cm \cite{hi4pi2016} and CO \cite{Dame2001} emission lines.  We include both components in our analysis from the map provided by the GALPROP code\footnote{\href{http://galprop.stanford.edu/}{galprop.stanford.edu}} \cite{Porter_2022}.  
We highlight that, based on our discussion in Ref.~\cite{Castro:2025wgf}, there exist a significant uncertainty regarding the adoption of gas template; as other column density maps, such as the one inferred by dust opacity measurements from the Planck Collaboration\footnote{\href{https://www.esa.int/Science_Exploration/Space_Science/Planck}{esa.int/Science$\_$Exploration/Space$\_$Science/Planck}} \cite{Planck:2016frx, Planck2011}, may suggest a reduction of gamma-ray fluxes of about $20-40\%$ in certain sky regions. To account for heavier elements in the gas, we scale the hydrogen density by a factor of 1.42, reflecting the Solar system composition, assumed to be representative of the entire Galactic disk \cite{Ferriere:2001rg}.

\par 
We estimate the total diffuse gamma-ray emission in the two sky regions probed by LHAASO ($|b|<5^\circ, 15^\circ<l<125^\circ$ and $125^\circ<l<235^\circ$) and in the two sky regions probed by Tibet-AS$\gamma$ experiment ($|b|<5^\circ, 25^\circ<l<100^\circ$ and $50^\circ<l<200^\circ$). We calculate the total Galactic neutrino flux and we compare it with the IceCube measurements.
%


%
%



\section{Results}
\label{sec:Results}
All the following results are reported for a Cygnus PeVatron of $W_{p}=2.8\times{10}^{51}$~erg and by considering a rescaling of the diffuse fluxes by the factors found in Ref.~\citep{Castro:2025wgf}, related to the lower gas column densities in the Galactic gas templates inferred by dust opacity measurements from Planck. This combination provides the best agreement within the multimessenger context considered in this work.

\subsection{Interpreting the CR knee with the Cygnus contribution}
\begin{figure}[t!]
\centering
\includegraphics[scale=0.292]{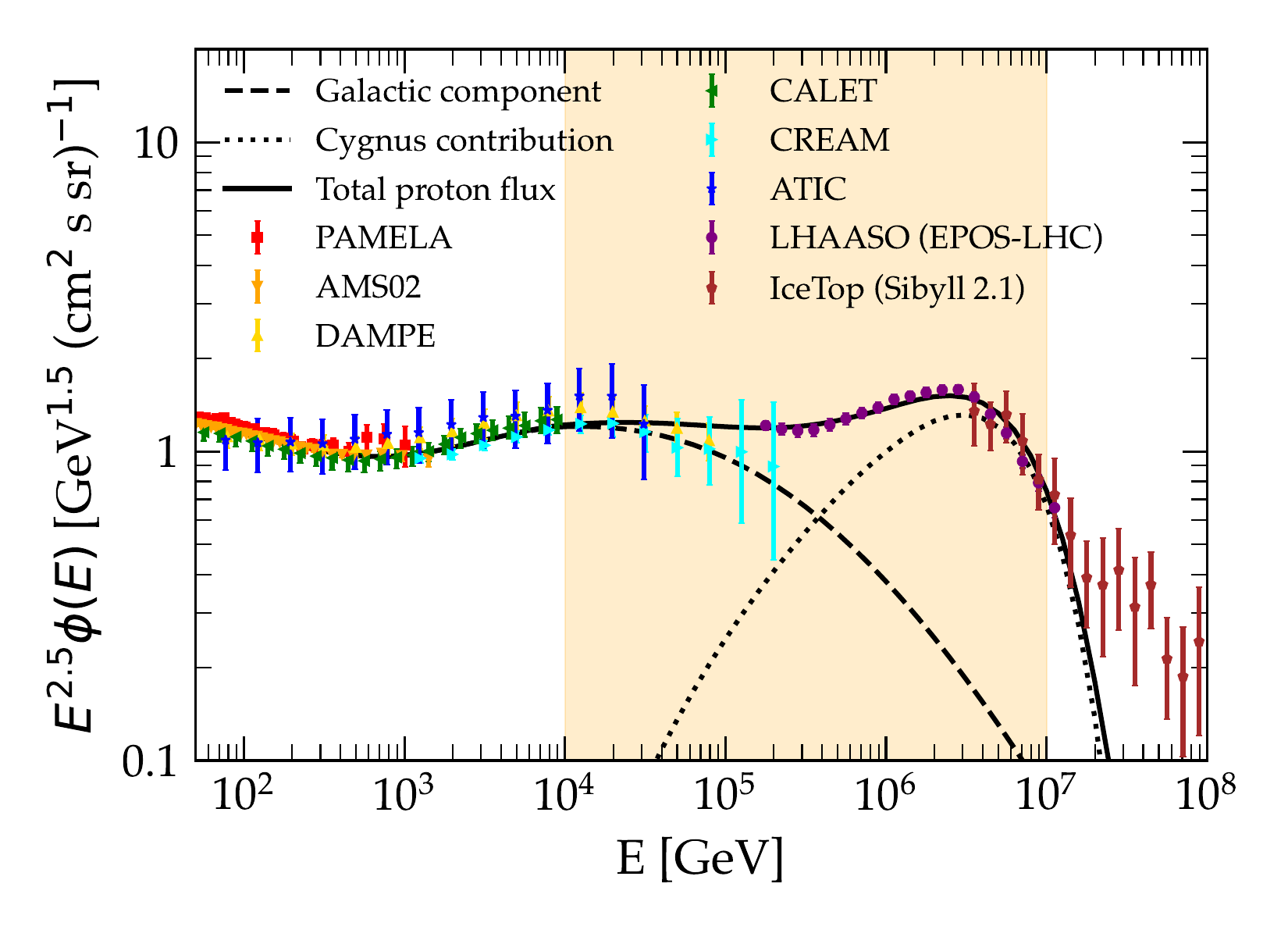}
\centering
\includegraphics[scale=0.292]{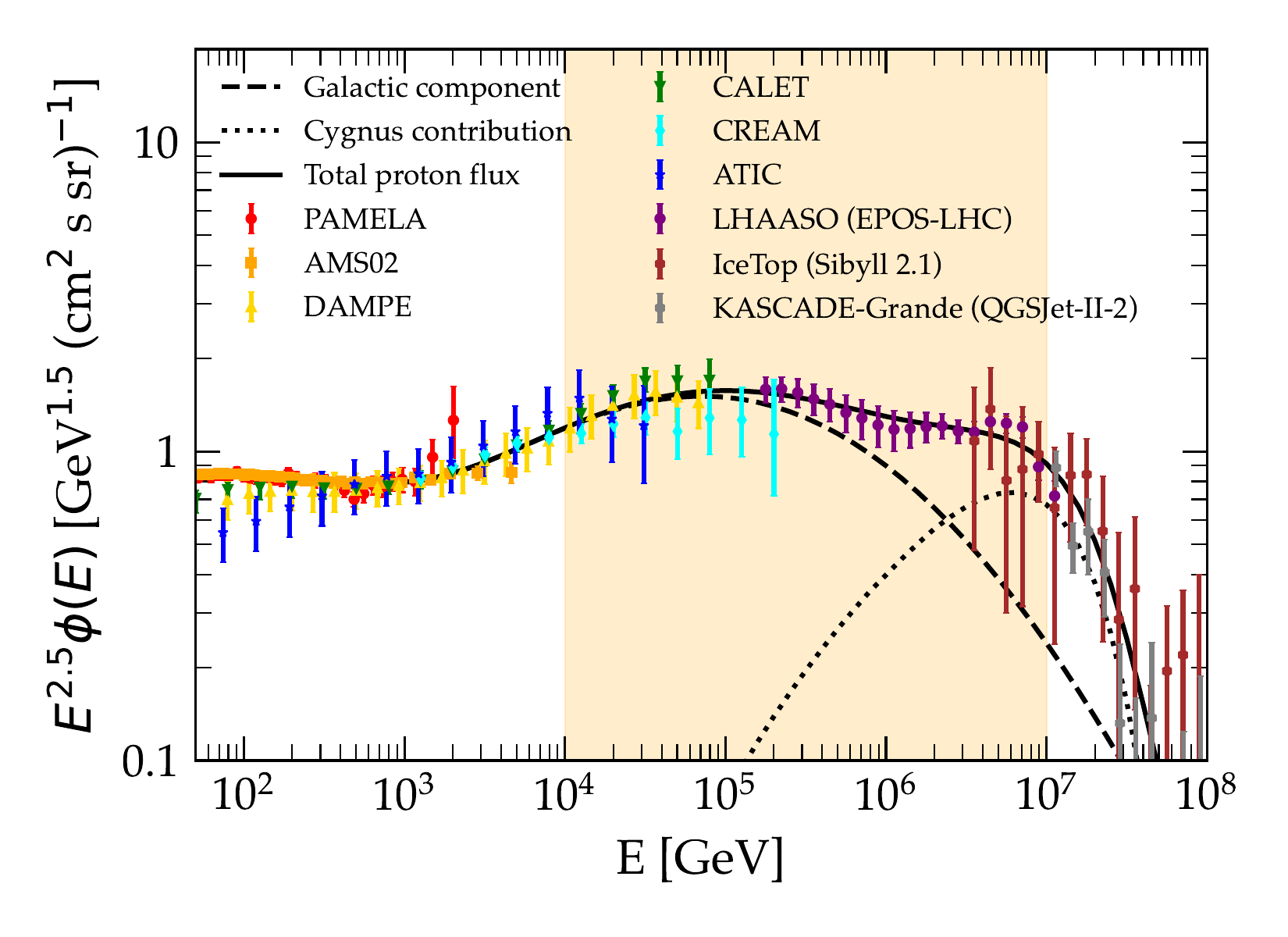}
\caption{Flux of CR protons (left panel) and helium (right panel) as a function of energy per nuclei. Observational data by direct detection experiments (PAMELA \cite{Adriani_2011}, AMS-02 \cite{Aguilar_2015a, Aguilar_2015b}, DAMPE \cite{An_2019, Alemanno_2021}, CALET \cite{Adriani_2019, Adriani_2023}, CREAM \cite{Yoon_2017} and ATIC-2 \cite{Panov_2009}) as well as ground-based observatories (IceTop \cite{Aartsen_2019}, KASCADE-Grande \cite{Finger_2011} and LHAASO \cite{Cao_2025b, LHAASO2025}) are shown with colored scatter points. Predicted Cygnus contributions are represented with black dotted-line for source injection parameters, $W_p=2.8\times10^{51}$~erg,  
$s=2.1$ and $E^{\rm max}=Z \cdot 5$~PeV, for atomic number $Z$. Best-fit of the Galactic population shown with black dashed-line. Black solid lines correspond to sum of both (Galactic and Cygnus) components and shaded areas indicate the energy range of interest for the production of TeV-PeV gamma rays and neutrinos.}
\label{fig:cosmic_ray_fluxes2}
\end{figure}

The predicted contribution of the Cygnus PeVatron to the local CR proton spectrum is shown in Fig.~\ref{fig:cosmic_ray_fluxes2} and compared with recent direct ~\citep{Adriani_2011, Aguilar_2015a, An_2019, Adriani_2019, Yoon_2017, Panov_2009} and ground-based measurements~\citep{LHAASO:2025byy, Aartsen_2019}. For the benchmark parameters adopted in this work (spectral index $s=2.1$; proton energy cutoff $E_p^{\rm max}=5$~PeV; proton luminosity $L_{p}=3\times 10^{37}$~erg/s; age $t_{\rm age}=3$~Myr), the Cygnus component provides a substantial contribution to the local flux at PeV energies. In particular, the model predicts that the source dominates the observed proton spectrum around $E_{\rm p}\sim 3-4$~PeV, naturally reproducing the knee feature measured by LHAASO.

We stress that our model parameters are not determined by fitting the CR data. We rather choose the parameters motivated by modeling the observations of the Cygnus region \cite{LHAASO:2023uhj}. Indeed, the assumed luminosity is a small fraction of the average kinetic power of collective stellar winds of Cygnus OB2 or powerful supernovae therein, the injection spectrum and suppressed diffusion in the Cygnus bubble are also inferred from LHAASO gamma-ray observations \cite{LHAASO:2023uhj}, while standard Kolmogorov diffusion is assumed in the outer region. In this framework the Cygnus contribution emerges as a natural consequence of the source properties inferred from gamma-ray observations.
Once the Cygnus contribution is fixed, the Galactic CR proton flux (dashed line in Fig. \ref{fig:cosmic_ray_fluxes2}) is obtained by requiring that the total flux (i.e., Galactic plus Cygnus components, solid line) reproduces observational data. We see that the Galactic component is subdominant above 300~TeV. 
%

Finally, in the right panel of Fig.~\ref{fig:cosmic_ray_fluxes2} we display the possible Cygnus contribution to the local $^{4}{\rm He}$ knee. We calculate this component by assuming that the $^{4}{\rm He}$ injection spectrum and diffusion coefficients scale with rigidity with respect to the proton case. The black dotted line in the right panel of Fig.~\ref{fig:cosmic_ray_fluxes2} corresponds to the prediction obtained by assuming that the total injected luminosity in helium (which is not fully constrained) is about $1/3$ of the proton luminosity. In this assumption, the Cygnus contribution explains the observed knee in the $^{4}{\rm He}$ spectrum at $\sim 6$~PeV, while the Galactic component has a substantial softening at few hundreds of TeV.

\subsection{Interpreting the Galactic diffuse gamma-ray and neutrino emission}
\begin{figure}[t!]
\centering
\includegraphics[scale=0.292]{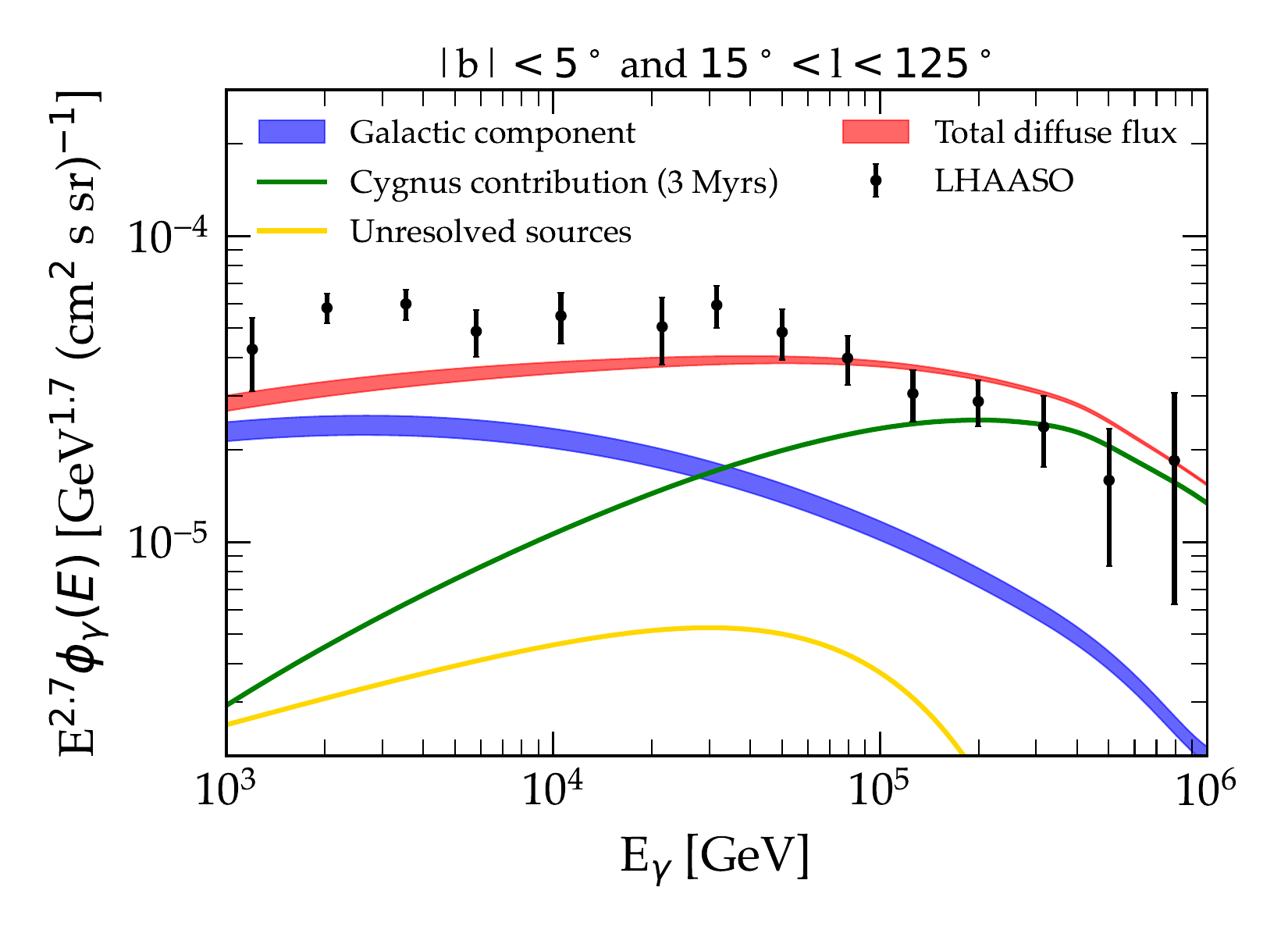}
\centering
\includegraphics[scale=0.292]{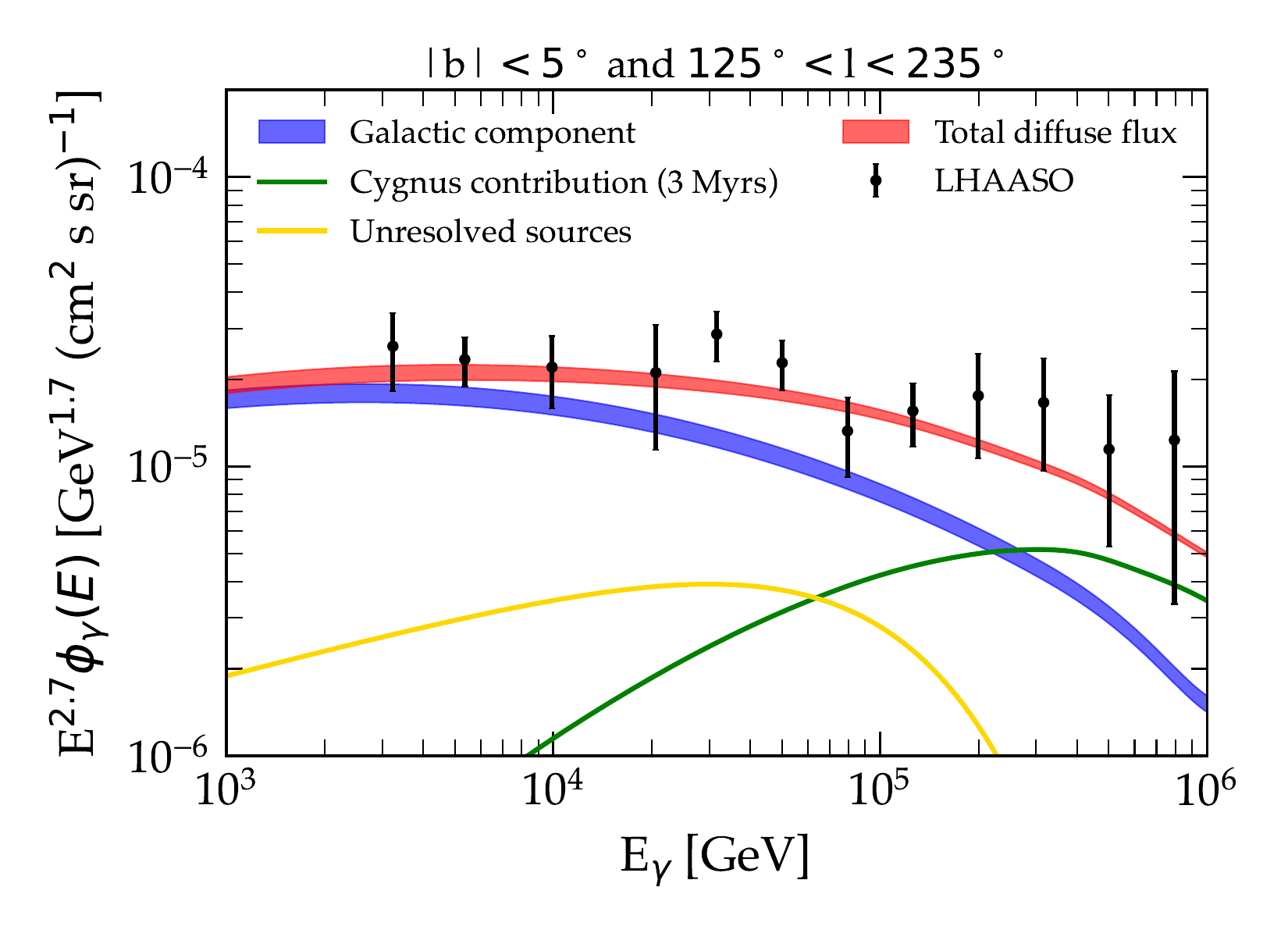}
\caption{Diffuse gamma-ray flux in LHAASO inner ($|b|<5^\circ$, $15^\circ<l<125^\circ$, top panel) and outer ($|b|<5^\circ$, $125^\circ<l<235^\circ$, bottom panel) Galaxy regions after LHAASO's mask is applied. The gamma-flux expectations for the Galactic component $\phi_\gamma^{\text{Gal}}(E_\gamma,\hat{n}_\gamma)$ the and Cygnus contribution $\phi_\gamma^{\text {Cyg}}(E_\gamma,\hat{n}_\gamma)$ are shown in blue bands and green lines, respectively. We also show with a yellow line, the contribution expected from unresolved sources in these regions as described in the text. The total diffuse gamma-ray flux predicted at Earth $\phi_\gamma (E_\gamma,\hat{n}_\gamma)$ is represented by red bands. Observational data by LHAASO \cite{Cao_2025a} are added with black scatter points.}
\label{fig:LHAASOmasked}
\end{figure}

The diffuse gamma-ray fluxes computed using the aforementioned methodology are presented in Fig.~\ref{fig:LHAASOmasked} for both sky regions probed by LHAASO and after applying the same mask used by the LHAASO Collaboration \cite{Cao_2025a}, intended to remove the contribution high-energy gamma-ray sources. Note that there are significant contributions from detected gamma-ray sources including pulsar wind nebulae (see Refs.~\cite{Fang:2023ffx,Vecchiotti:2024kkz} and Supplementary Material for the gamma-ray emission without the mask). 
The diffuse flux originating from the Cygnus region is shown with green lines, while the flux produced by the Galactic CR population is shown with blue bands. Here we also include the average contribution expected by unresolved pulsar wind nebulae following Ref.\cite{Vecchiotti:2024kkz} (for details see \ref{App:Unresolved}). The sum of all components represents the total diffuse gamma-ray flux and is presented with red lines. We find an agreement with the LHAASO data in both regions and a marginal role of the unresolved sources, as expected due to the masking procedure \cite{Vecchiotti:2024kkz}. 
\par
\begin{figure}[t!]
\centering
\includegraphics[scale=0.3]{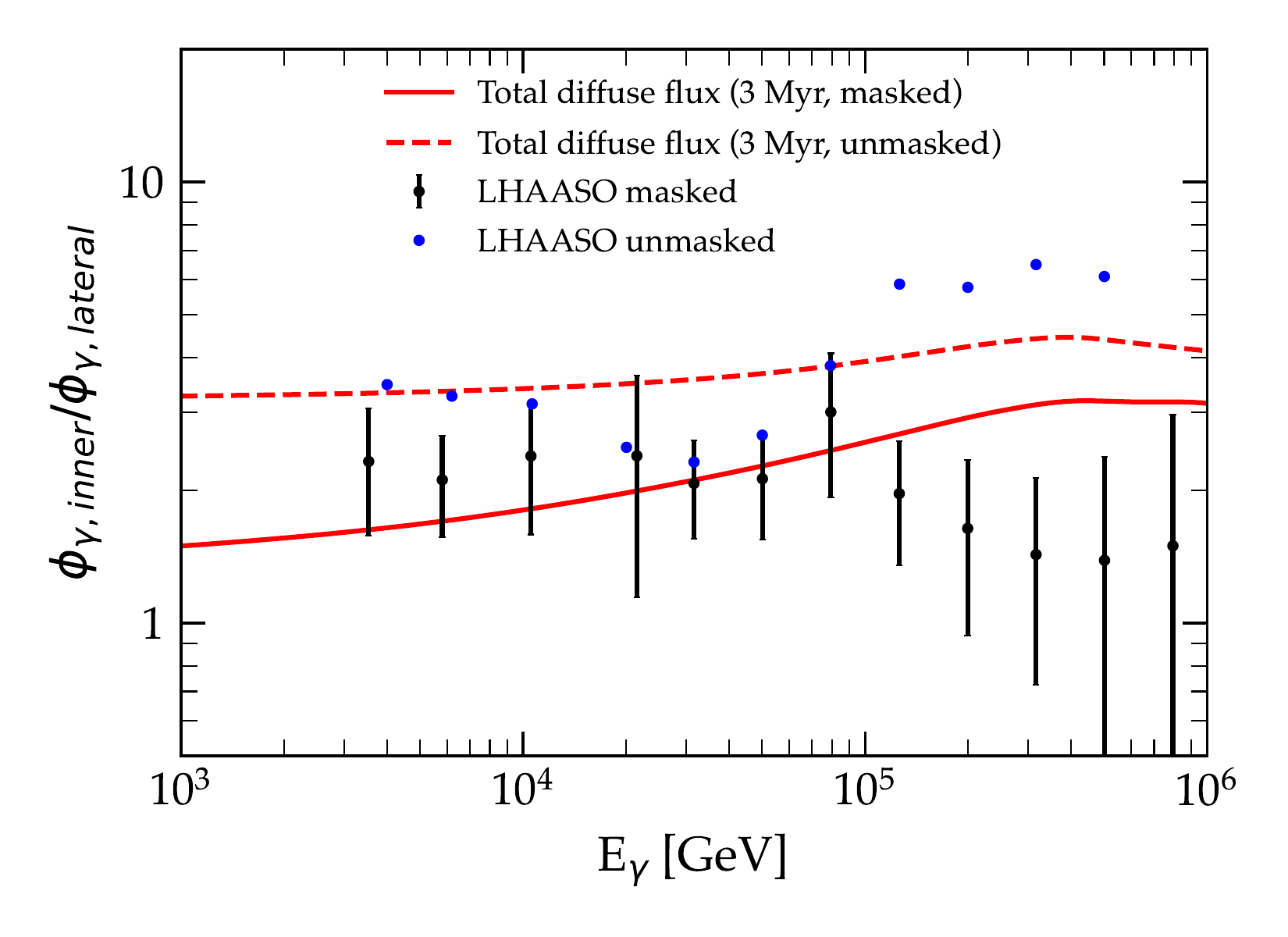}
\caption{ Ratio of diffuse gamma-ray flux between LHAASO inner and outer Galaxy regions. The ratio between LHAASO observation data after the mask is applied is shown with black scatter points, while the observations before the mask are shown with blue scatter points. The ratio from our computed diffuse fluxes are shown with red solid and dashed lines, for masked and unmasked cases, respectively.}
\label{fig:ratio}
\end{figure}

On the other hand the presence of the CR knee changes the spectral shape of the total expected diffuse gamma-ray emission in both sky regions, which may generate the apparent tension between CRs and diffuse gamma-ray discussed in Ref.~\cite{Castro:2025wgf}. However, in our scenario, this is interpreted as an effect of the mask. This is reflected in Fig.~\ref{fig:ratio}, where we plot the ratio between the gamma-ray fluxes in the inner and outer Galaxy regions. When the mask is removed, the entire contribution of the Cygnus region appears in the inner observational window, and the ratio between the total gamma-ray emission data (blue points) in both regions shows a ``bump"-like feature at energies $E_{\gamma}>100$ TeV, compatible with the presence of an additional source contribution to the CR knee mainly in the inner region. This prediction for the ratio, obtained from our computed fluxes under the same hypothesis of unmasked regions (i.e. those of Fig.~\ref{fig:LHAASOunmasked}), is shown with a red dashed line.
\par 
In Fig.~\ref{fig:ratio} the black data points represent the ratio between the inner and outer LHAASO regions with the mask applied by the LHAASO Collaboration. This mask in the inner region excludes a large portion of the sky around the Cygnus region, thus removing not only LHAASO's bubble region but also a part of the external region outside the bubble.
This fact not only explains the absence of a structure corresponding to the CR knee in Fig.~\ref{fig:LHAASOunmasked} but also makes the ratio more ``flat'' in energy in Fig.~\ref{fig:ratio}, when we compare the data to our prediction (corresponding to the red solid line) based on the computed fluxes. A slight discrepancy remains at energies above 100 TeV, however the uncertainty in the Galactic gas contained in the Cygnus region leads to large uncertainty in the computation of the ratio at those energies once the mask is applied, especially in the small sky region covered by the Cygnus region in the outer Galaxy region. Small variations of the Galactic gas column density in this region can significantly make the bump-like feature in the ratio more or less pronounced. Despite these limitations, the recovery of a spectral structure at $E_\gamma\sim(200-300)$~TeV, corresponding to the CR proton knee energy, is suggestive and reinforces the importance of a nearby PeVatron in the inner Galactic region.

\begin{figure}[t!]
\centering
\includegraphics[scale=0.292]{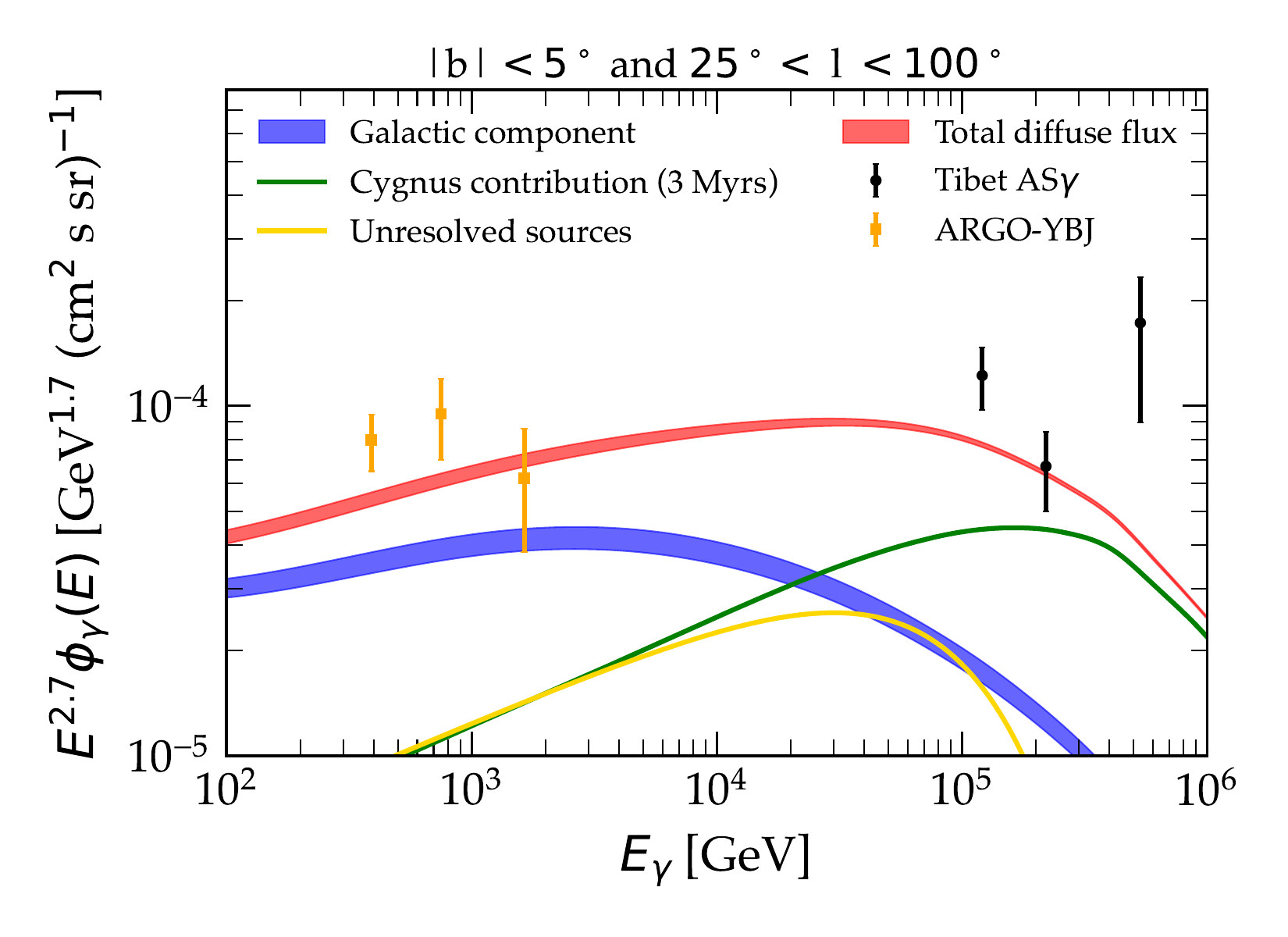}
\centering
\includegraphics[scale=0.292]{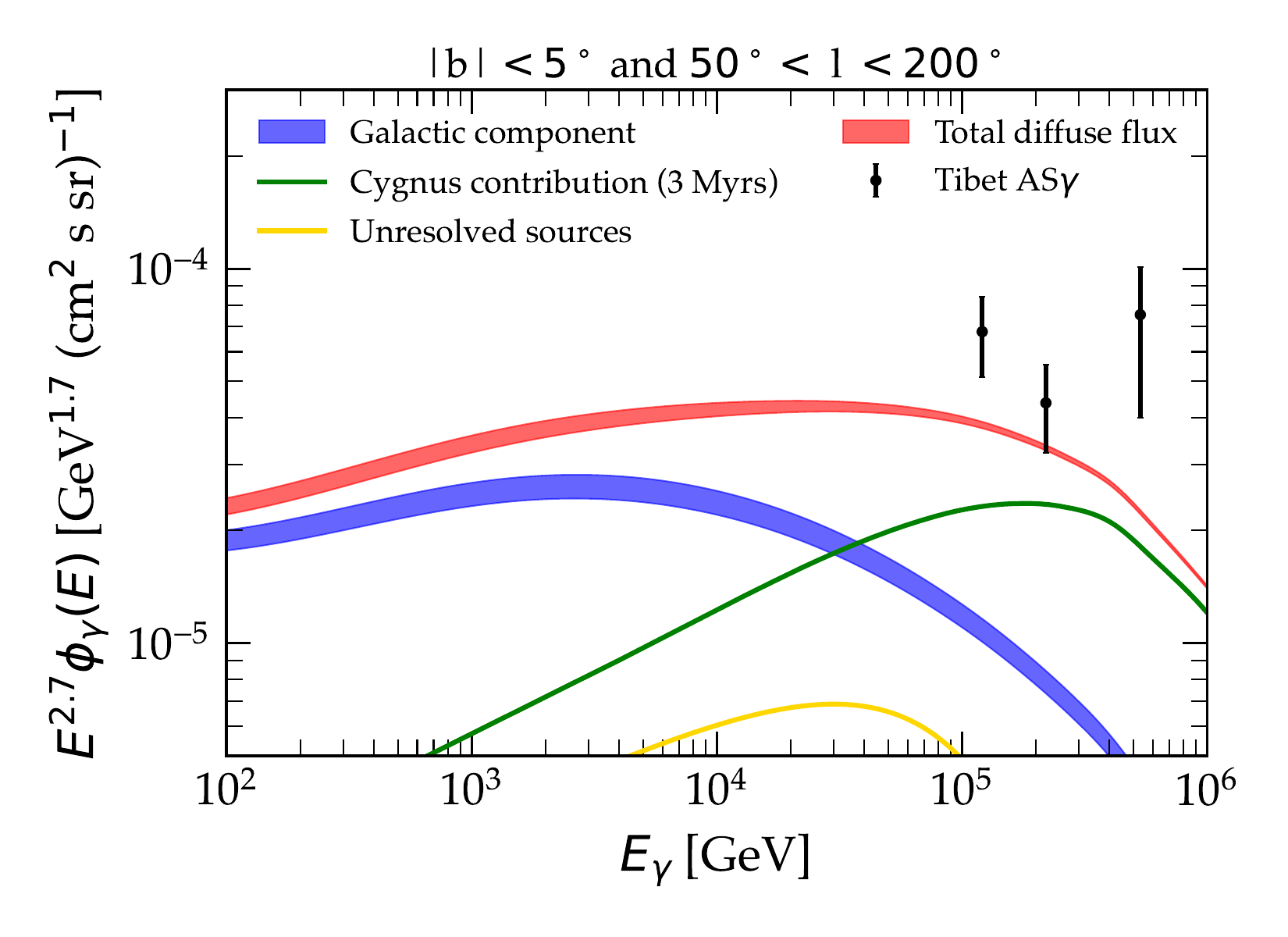}
\caption{Diffuse gamma-ray flux in Tibet-AS$\gamma$ inner ($|b|<5^\circ$, $15^\circ<l<125^\circ$, top panel) and lateral ($|b|<5^\circ$, $125^\circ<l<235^\circ$, bottom panel) Galaxy regions. The gamma-flux expectations for the Galactic component $\phi_\gamma^{\text{Gal}}(E_\gamma,\hat{n}_\gamma)$ the and Cygnus contribution $\phi_\gamma^{\text {Cyg}}(E_\gamma,\hat{n}_\gamma)$ are shown in blue bands and green lines, respectively. We add the contribution expected from unresolved sources in these regions with yellow lines. The total diffuse gamma-ray flux predicted at Earth $\phi_\gamma (E_\gamma,\hat{n}_\gamma)$ is represented by red bands. Observational data by Tibet-AS$\gamma$ \cite{Amenomori_2021,Kato:2024ybi} and ARGO-YBJ \cite{Bartoli_2015} are added with black and orange scatter points, respectively.}
\label{fig:Tibet}
\end{figure}

\par
The Galactic diffuse gamma-ray fluxes computed for both sky regions probed by Tibet-AS$\gamma$ experiment are presented in Fig.~\ref{fig:Tibet}. The Tibet-AS$\gamma$ data are scaled in order to subtract the contamination due to high-energy sources observed by LHAASO following Ref.\cite{Kato:2024ybi}. In this case, the contribution of unresolved sources (yellow line) is not negligible and needs to be accounted for in order to have general agreement with the observations, as discussed in Refs.~\cite{Fang:2021ylv,Vecchiotti:2021yzk}. However, in the energy range of the Tibet-AS$\gamma$ data points, the contribution of the Cygnus PeVatron dominates the total diffuse emission. As has been discussed in Section \ref{sec:intro}, if the gamma-ray flux observations of Tibet-AS$\gamma$ experiment are rescaled by the gas column density in the two sky regions, one finds an overlap between both datasets. If the Cygnus contribution is dominant in both measurements as we have found, this can consistently explain this observational aspect.
%

%
\begin{figure}[t!]
\centering
\includegraphics[scale=0.3]{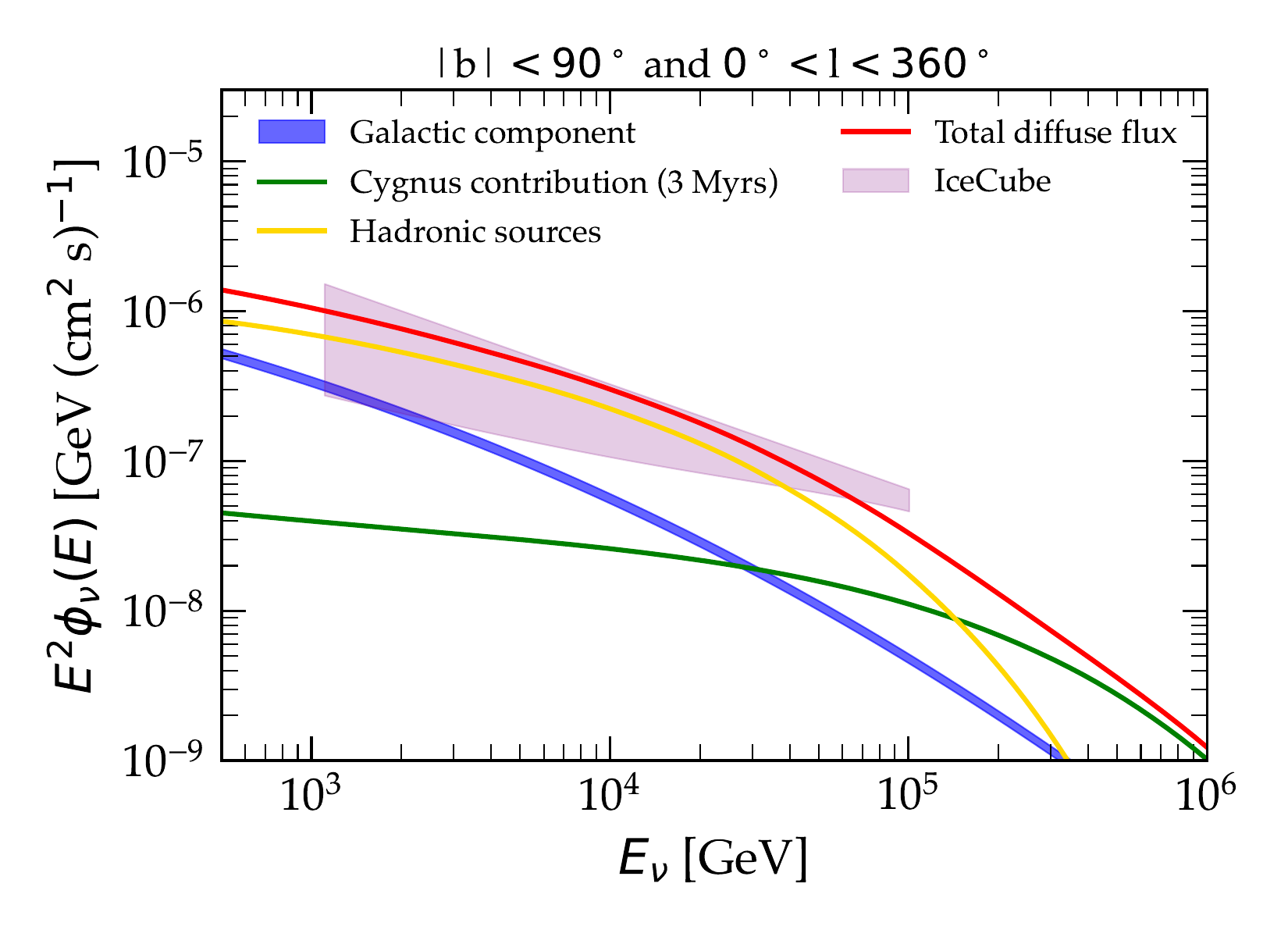}
\caption{Galactic diffuse (all-flavor) neutrino flux from the Galactic Plane ($|b|<90^\circ$, $0^\circ<l<360^\circ$). The neutrino flux expectations for the Galactic component $\phi_\nu^{\text{Gal}}(E_\nu,\hat{n}_\nu)$ the and Cygnus contribution $\phi_\nu^{\text {Cyg}}(E_\nu,\hat{n}_\nu)$ are shown in a blue band and green line, respectively. The total diffuse gamma-ray flux predicted at Earth $\phi_\nu (E_\nu,\hat{n}_\nu)$ is represented by a red band. We add the contribution expected from unresolved sources with a yellow line, assuming 40$\%$ of the Galactic gamma-ray sources are hadronic. Best-fit of the diffuse flux based on IceCube observational data \cite{IceCube:2023ame} is represented with purple shaded area.}
\label{fig:neutrinos}
\end{figure}
%
Finally, we estimate the neutrino counterpart of the total Galactic emission when the Cygnus PeVatron contribution is included. 
Results are reported in Fig.~\ref{fig:neutrinos} and are compared with the IceCube measurements. In particular, IceCube measurements of the Galactic emission are based on three different templates ($\pi^0$, KRA$^5$ and KRA$^{50}$ models \citep{Gaggero_2015}) and are sensitive to an energy range of $E_\nu\sim 1-100$~TeV. We show with a shaded magenta band the uncertainty on the measurements due to the former two different templates adopted, corresponding to the area spanned by the maximum and minimum fluxes from the templates. Our prediction for the Cygnus contribution is reported with a green line, while the Galactic contribution due the CR sea is reported with a blue band and includes also heavy elements contribution as discussed in Section~\ref{sec:Methods}. As for the diffuse gamma-ray emission, also in the neutrino side the Cygnus contribution will dominates over the Galactic diffuse emission at energies above $E_\nu \gtrsim 30$ TeV.  

However, in the neutrino sky, hadronic Galactic sources are also expected to contribute to the diffuse emission measured by IceCube being all unresolved \cite{Ahlers:2013xia,Fang:2021ylv,Fang:2023ffx}. We then include the contribution expected for the Galactic population following the approach we described in Ref.~\citep{Vecchiotti:2023ill}. Namely, we estimate the total Galactic population of high-energy gamma-ray sources with a population study tuned to reproduce all the characteristics of the HGPS sources catalog, a similar study to the one providing the contribution of unresolved gamma-ray sources in the previous plots. We then include the fraction of the total cumulative flux due to the Galactic sources allowed within the IceCube measurements uncertainty. We find that, for a cutoff energy of 500 TeV of the sources, this fraction is the $40\%$ of the total flux and is reported with a yellow line in Fig.~\ref{fig:neutrinos}.    

The Cygnus region has been discussed for many years as a promising source of high-energy neutrinos, especially after Milagro's discovery of extended TeV gamma-ray emission \cite{Beacom:2007yu,Halzen:2007ah}. Later observations of the Cygnus cocoon and the larger Cygnus bubble by Fermi-LAT \cite{Fermi:2011}, HAWC \cite{Abeysekara:2021yum}, and LHAASO \cite{LHAASO:2023uhj} motivated increasingly detailed calculations of the neutrino flux associated with the observed gamma-ray emission \cite{Yoast-Hull:2017gaj,Fang:2021ylv,Li:2024gnb}. The key distinction of our model is that the Cygnus region is not treated simply as another gamma-ray-bright neutrino source. Instead, we argue that it contributes a hard spectral component in the knee region of the Galactic CR proton spectrum. This leads to a distinct multimessenger prediction: the gamma-ray emission observed from the Cygnus cocoon/bubble and the diffuse emission measured by Tibet AS$\gamma$/LHAASO are interpreted as manifestations of the same underlying knee-producing population, with a neutrino counterpart that may be challenging for current IceCube data but within the reach of IceCube-Gen2.

\section{Conclusion}
We investigated the possibility that the CR knee arises from the contribution of a nearby Galactic PeVatron located in the Cygnus region. Using a physically motivated two-zone diffusion model and source parameters consistent with LHAASO observations of the Cygnus bubble, we constructed a self-consistent multimessenger scenario connecting CR, gamma-ray and neutrino data.

Our results show that a Cygnus PeVatron, with a total proton energy of $W_{p}\sim3\times{10}^{51}$~erg, reproduces the proton knee observed at Earth while remaining compatible with existing gamma-ray and neutrino measurements. In this picture, the local CR spectrum at PeV energies is dominated by the contribution of a nearby accelerator rather than by the large-scale Galactic CR population, which is expected to cutoff at few hundreds of TeV. 

The observed CR anisotropy in the PeV range would not provide a robust exclusion of a significant Cygnus contribution. In general, the dipole amplitude scales as $\delta\sim(3D/c)|\nabla n|/n$. For an extended and possibly multiepisodic source complex such as Cygnus, the expected gradient depends sensitively on the effective source age, spatial extent, and anisotropic transport considering details of the magnetic field structure. Consequently, simple single-source burst-like estimates would be interpreted as upper limits, and the current $\sim0.1$\% level anisotropy would not be in obvious conflict with a substantial Cygnus contribution.

Our model also explains several observational features of the very-high-energy diffuse gamma-ray sky, including the enhancement of the emission in the inner Galaxy and the absence of a gamma-ray spectral break corresponding to the CR proton knee. At energies above $\sim 100$ TeV, 
the extended emission produced by the Cygnus region becomes a dominant component of the diffuse gamma-ray and neutrino fluxes. 

Taken together, these results provide a concordance multimessenger interpretation linking the Cygnus PeVatron to the origin of the CR knee. Future measurements of Galactic diffuse gamma rays and neutrinos at PeV energies will offer a decisive test of this scenario and of the role of nearby PeVatrons in shaping the high-energy CR spectrum.

\begin{acknowledgments}
The work of K.M. was supported by the NSF Grants No.~AST-1908689, No.~AST-2108466 and No.~AST-2108467, and KAKENHI No.~20H01901 and No.~20H05852.
The work of V. V. is supported by the European Union’s Horizon Europe research and innovation programme under the Marie Skłodowska-Curie grant agreement No. 101208655 (CORNO GRANDE–COnstRaiNing the Origin of Galactic cosmic RAys using gamma-ray and Neutrino Diffuse Emissions).

Recently, while this work was being completed, Ref.~\citep{Fang:2026ydz} explored a related scenario arguing that a nearby source could be consistent with the flux of CR protons and CR anisotropy at PeV energies.

\end{acknowledgments}

\appendix

\section{Galactic Gamma-Ray Flux without the Mask}
\label{App:GalacticUnmasked}
\begin{figure}[t!]
\centering
\includegraphics[scale=0.292]{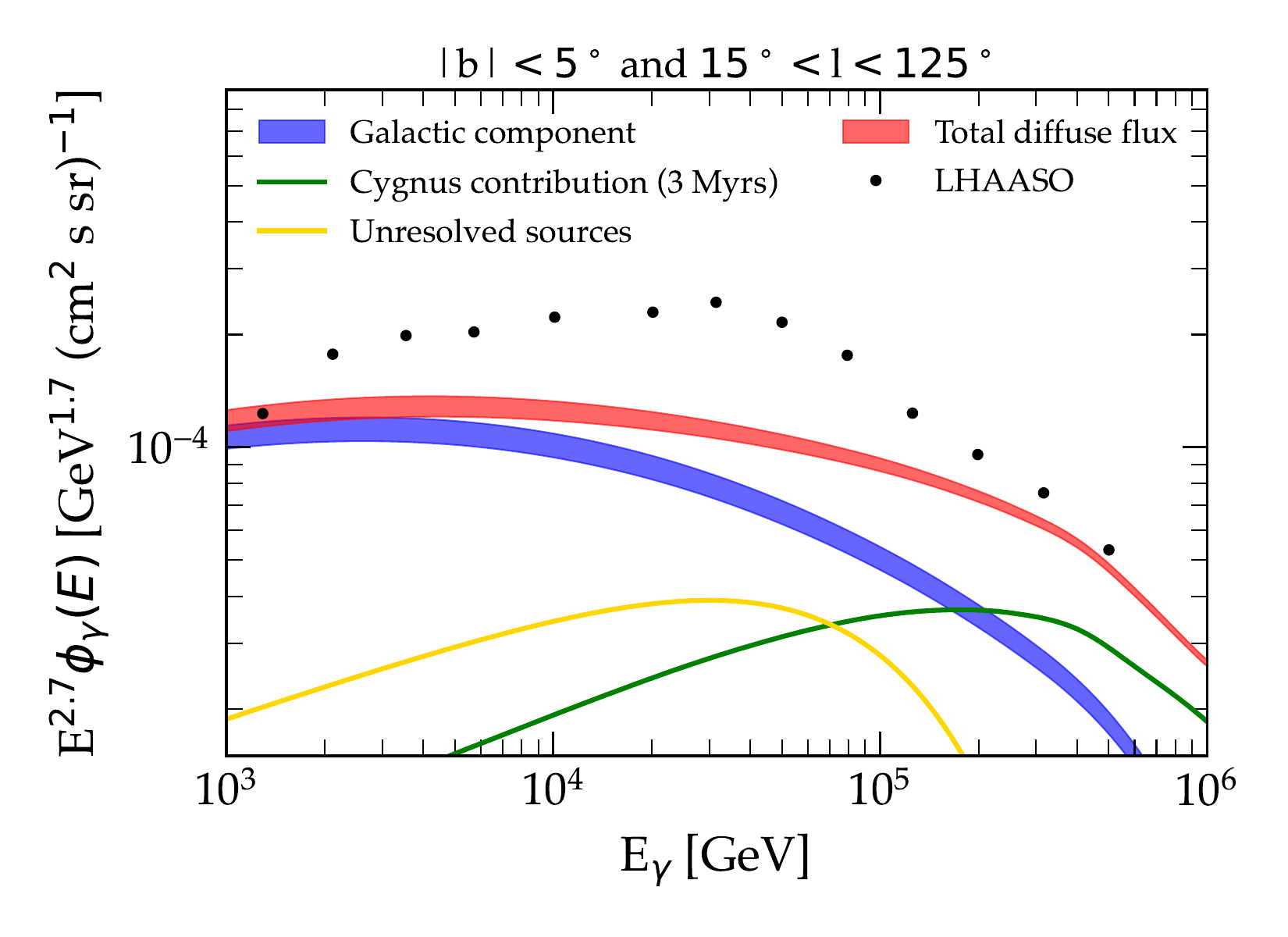}
\centering
\includegraphics[scale=0.292]{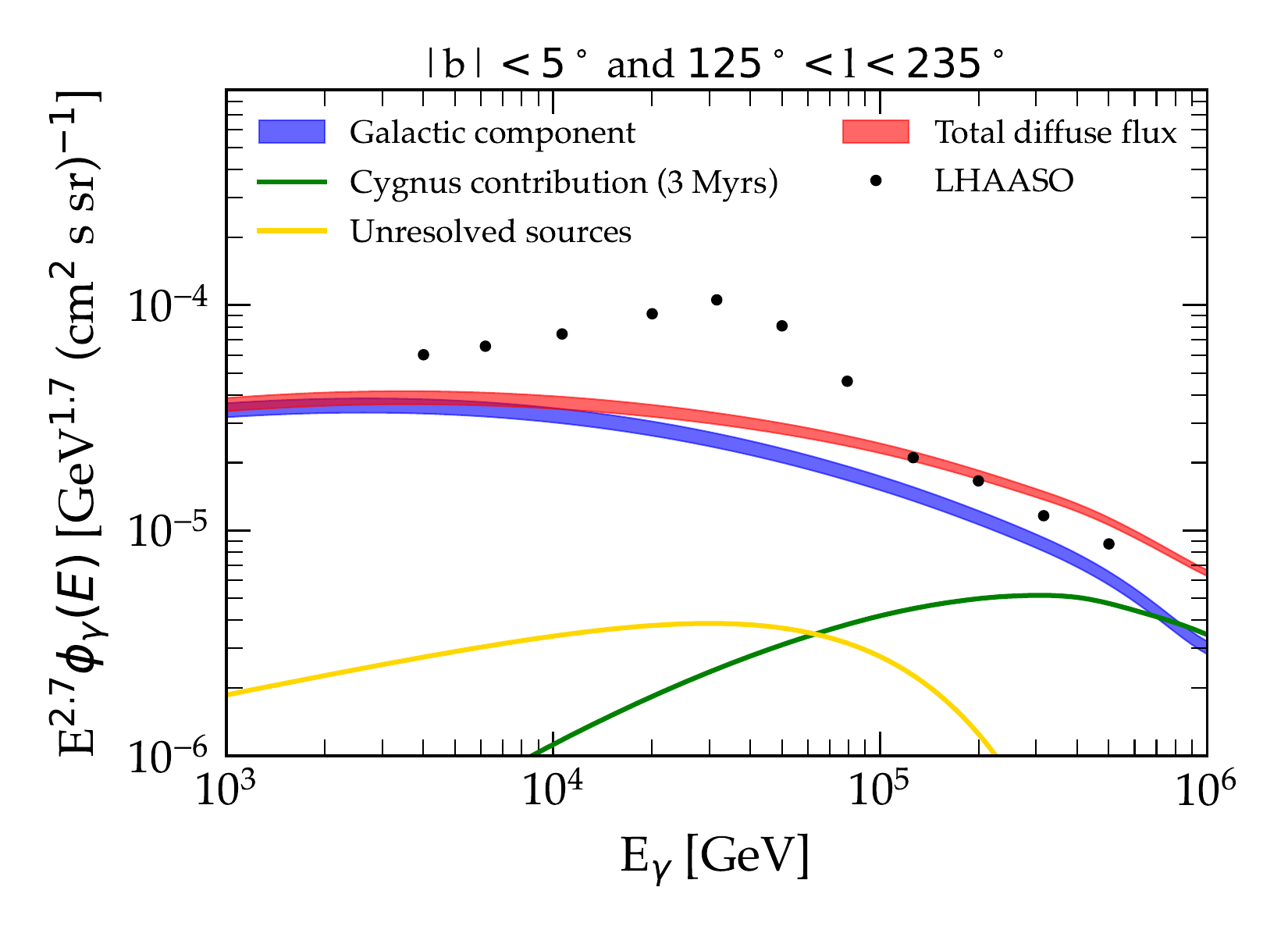}
\caption{The gamma-ray flux in LHAASO inner ($|b|<5^\circ$, $15^\circ<l<125^\circ$, top panel) and outer ($|b|<5^\circ$, $125^\circ<l<235^\circ$, bottom panel) Galaxy regions without the mask. The gamma-flux expectations for the Galactic component $\phi_\gamma^{\text{Gal}}(E_\gamma,\hat{n}_\gamma)$ the and Cygnus contribution $\phi_\gamma^{\text {Cyg}}(E_\gamma,\hat{n}_\gamma)$ are shown in blue bands and green lines, respectively. We also show with a yellow line, the contribution expected from unresolved sources in these regions as described in the text. The total diffuse gamma-ray flux predicted at Earth $\phi_\gamma (E_\gamma,\hat{n}_\gamma)$ is represented by red bands. Observational data by LHAASO \cite{Cao_2025a} are added with black scatter points, without error bars.}
\label{fig:LHAASOunmasked}
\end{figure}
In Fig.~\ref{fig:LHAASOunmasked} we show our expectations for the LHAASO regions once that the mask is removed. The color code is the same of Fig.~\ref{fig:LHAASOmasked}. The black scattered data points (for which we are unable to add error bars) represent the total gamma-ray emission measured by LHAASO in these regions, including the contribution from resolved gamma-ray sources. In this case, our computed diffuse fluxes (the red lines representative) are required to stay below the data points, as they correspond to the pure diffuse emission. At energy $E_{\gamma}>200$ TeV the contribution of the Cygnus PeVatron dominates the total emission in the inner Galactic region. 
\par 

\section{Unresolved Pulsar Wind Nebulae}
\label{App:Unresolved}

The contribution of unresolved sources to the LHAASO and Tibet-AS$\gamma$ signals is calculated following Ref.~\cite{Vecchiotti:2024kkz}. In that work, a synthetic population of PWNe is generated. The intrinsic luminosity of these sources, $L$ (integrated in the $1-100$ TeV energy range), is assumed to decrease with the source age $t_{\rm age}$ according to
\begin{equation}
L(t_{\rm age})= L_{\rm max}\left(1+\frac{t_{\rm age}}{\tau_{\rm sd}}\right)^{-2},
\label{lum}
\end{equation}
where $L_{\rm max}$ is the maximum luminosity and $\tau_{\rm sd}$ is the pulsar spin-down timescale. The source age is uniformly sampled in the interval $[0,\,T]$, where $T=10^6$~yr represents the assumed duration of very-high-energy emission. The sources are spatially distributed according to Ref.~\cite{Lorimer:2006qs}, and follow an exponential distribution along the direction $z$ perpendicular to the Galactic plane, $\propto \exp(-|z|/H)$, where $H$ represents the thickness of the Galactic disk.

The best-fit values used for the calculation of the LHAASO sources are $L_{\rm max} = 1.87 \times 10^{35}\,\rm erg\,s^{-1}$ and $\tau_{\rm sd}=2.9$ kyr \citep{Vecchiotti:2024kkz}. These values are obtained fitting the H.E.S.S. Galactic Plane Survey data assuming a Galactic disk thickness $H=0.05\,$kpc (see,e.g., \cite{Cataldo:2020qla}). The best-fit values are derived under the assumption that all sources have a power-law spectrum with spectral index $2.4$, corresponding to the average spectral index of the brightest sources observed by H.E.S.S., and an exponential cut-off at $E_{\rm max}=100$ TeV. 

To calculate the unresolved source contribution, the KM2A and WCDA sensitivity thresholds for point-like sources with spectral index $3$ and $2.5$, respectively, are adopted, as reported by the LHAASO collaboration in Fig.~5 of Ref.~\cite{LHAASO:2023rpg}. 
A source is considered unresolved if it is simultaneously unresolved by both LHAASO detectors.
All sources are considered point-like; if the sources were extended, their contribution would be larger (see \cite{Vecchiotti:2024kkz}).

We use the same best-fit values to calculate the unresolved source contribution in the Tibet-AS$\gamma$ regions, adopting only the KM2A detection threshold. This choice is motivated by the fact that the Tibet-AS$\gamma$ data reported in Fig.~\ref{fig:Tibet} are rescaled to account for contamination from LHAASO resolved sources \cite{Kato:2024ybi}. These assumptions slightly differ from those adopted in Ref.~\citep{Vecchiotti:2021yzk}. In particular, the thickness of the disk $H=0.05$~kpc instead of $H=0.2$~kpc, and the assumed cut-off energy of $100$~TeV rather than the $500$~TeV used in the previous work to increase compatibility with LHAASO measurements. However, the main difference arises from the different assumed detection threshold.

\bibliography{biblio}
\end{document}